\documentclass[epj,twocolumn]{webofc}
\usepackage[varg]{txfonts}   
\usepackage{color}
\woctitle{Powders \& Grains 2017}

\newcommand{\be}{\begin{equation}}
\newcommand{\ee}{\end{equation}}
\newcommand{\ba}{\begin{eqnarray}}
\newcommand{\ea}{\end{eqnarray}}

\definecolor{red}{rgb}{1,0,0}
\definecolor{blue}{rgb}{0,0,1}
\definecolor{black}{rgb}{0,0,0}

\def\Ic{I}
\def\Jc{J}
\def\Lc{L}

\def\Pc{{\cal P}}
\def\edot{\dot\epsilon}

\newcommand{\eq}[1]{\begin{align}#1\end{align}}

\def\fv{\vec{f}}
\def\Fv{\vec{F}}
\def\Vv{\vec{V}}

\def\Uv{\vec{U}}
\def\Dc{{\cal D}}

\def\Pc{{\cal P}}

\begin{document}
\title{Unifying Suspension and Granular flows near Jamming}
%
%

\author{\firstname{Eric} \lastname{DeGiuli}\inst{1,3}\fnsep\thanks{\email{eric.degiuli@epfl.ch}} \and
        \firstname{Matthieu} \lastname{Wyart}\inst{2}\fnsep\thanks{\email{matthieu.wyart@epfl.ch}} 
}

\institute{ Institute of Physics, \'Ecole Polytechnique F\'ed\'erale de Lausanne (EPFL), CH-1015 Lausanne, Switzerland
          }

\abstract{%
Rheological properties of dense flows of hard particles are singular as one approaches the jamming threshold where flow ceases, both for granular flows dominated by inertia, and for over-damped suspensions. Concomitantly, the lengthscale characterizing velocity correlations appears to diverge at jamming.  Here we review a theoretical framework that gives a scaling description of stationary flows of frictionless particles.  Our analysis applies {\it both} to suspensions and inertial flows of hard particles. We report numerical results in support of the theory, and show the phase diagram that results when friction is added, delineating the regime of validity of the frictionless theory.
}
\maketitle
\section{Introduction}
 
Microscopic description of particulate materials such as grains, emulsions or suspensions is
complicated by the presence of disorder, and by the fact that these systems are often out-of-equilibrium.
One of the most vexing problems is how these materials transition between a flowing and a solid phase.  
When this transition is driven by temperature, it corresponds  to the glass transition where a liquid becomes a glass, an amorphous structure that cannot flow on experimental time scales. Here we focus instead on athermal systems driven by an imposed stress, such as granular flows, and consider both the case where inertia is important (such as in aerial granular flows) or not (such as over-damped suspensions). We focus primarily on the case of hard particles. 

 Empirical constitutive relations have been proposed to describe such dense flows in the limit of hard particles \cite{MiDi04,Cruz05,Jop06}. Two important dimensionless quantities are the packing fraction $\phi$ and the stress ratio $\mu\equiv \sigma/p$  (also called the effective friction), where $\sigma$ is the applied shear stress and $p$ the pressure carried by the particles. For inertial flow, dimensional analysis implies that both quantities can only depend on the strain rate $\dot\epsilon$, $p$, the particle diameter $D$ and the mass density of the hard particles $\rho$ via the inertial number $\Ic\equiv \dot \epsilon D \sqrt{\rho/p}$. One finds empirically that the constitutive relations $\mu(\Ic)$ and $\phi(\Ic)$ converge to a constant as $\Ic\rightarrow 0$, corresponding to the jamming transition where flow stops. We define $\mu(0)\equiv \mu_c$ and $\phi(0)\equiv\phi_c$, which are system-specific and depend on particle shape, poly-dispersity, friction coefficient, etc. 
 Near jamming, the constitutive relations are observed to be singular with:
\ba
\label{0}
\delta \mu & \equiv \mu(\Ic)-\mu_c\propto \Ic^{\alpha_\mu} \\
\delta \phi &\equiv \phi_c-\phi(\Ic)\propto \Ic^{\alpha_\phi} \label{01}
\ea
As jamming is approached the dynamics becomes increasingly  correlated in space \cite{Pouliquen04,Olsson07}. By considering the dominant decay \cite{During14} of the velocity correlation function, one can define a length scale $\ell_c$: 
\be
\label{02}
\ell_c \sim \Ic^{-\alpha_\ell}
\ee
Similar dimensional arguments have been made for dense suspensions of non-Brownian particles  \cite{Lemaitre09b,Boyer11}. 
In that case the  relevant dimensionless number is the viscous number $\Jc= \eta_0\dot\epsilon/p$, where $\eta_0$ is the viscosity of the solvent. Empirically one finds similar relations:
\ba
 \label{03}
\delta \mu & \equiv \mu(\Jc)-\mu_c\propto \Jc^{\gamma_\mu} \\
\delta \phi& \equiv \phi_c-\phi(\Jc)\propto \Jc^{\gamma_\phi} \label{04}\\
\ell_c & \sim \Jc^{-\gamma_\ell}  \label{05}
\ea
These relations imply that the viscosity  $\eta=\sigma/\edot$ of the suspension diverges as jamming is approached. Indeed Eq.(\ref{03}) implies that $\sigma\sim p$ near jamming (in our scaling arguments below we may thus exchange freely $\sigma$ and $p$), so that $\Jc\propto \eta_0/\eta$. Eq.(\ref{04}) then implies that:
\be
\label{06}
\frac{\eta}{\eta_0}\propto (\phi_c-\phi)^{-1/\gamma_\phi}
\ee
 When both viscosity and inertia are present, a transition from viscous to inertial flow occurs as strain rate $\edot$ is increased at fixed volume fraction \cite{Fall10,Trulsson12,Vagberg14}. This defines a cross-over strain rate
\eq{
\label{epsco}
\edot_{v\rightarrow i} \propto \frac{\eta_0}{\rho D^{2}} (\phi_c - \phi)^{\gamma_{\edot}},
}
where the prefactors follow from a dimensional analysis.

\begin{figure}[bh!] 
\includegraphics[width=0.50\textwidth,clip]{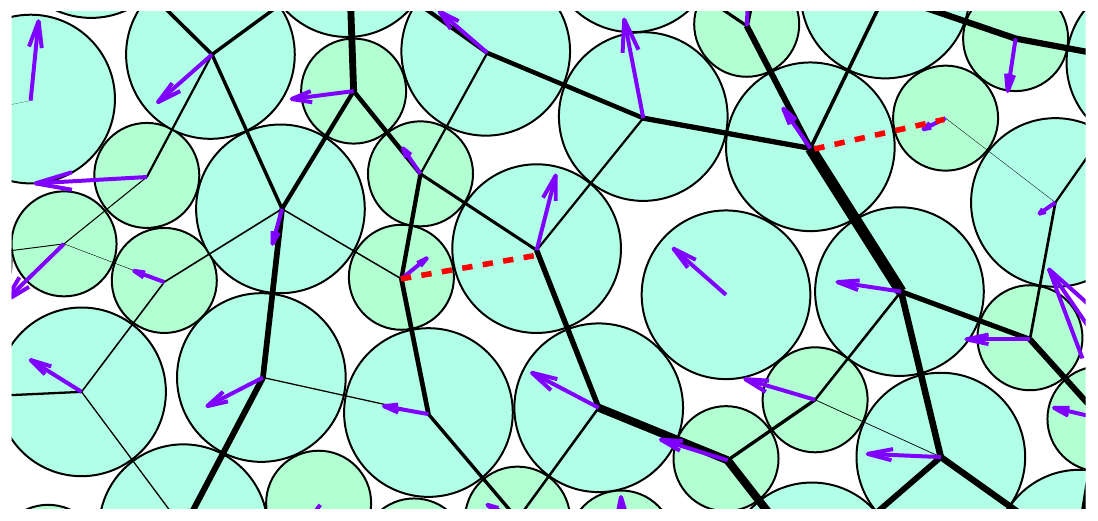}
\caption{Illustration of solid destabilization: several weak contacts, indicated by red dashed lines, are opened. This induces a space of extended, disordered floppy modes, one of which is shown (arrows). Line thickness indicates force magnitude in the original, stable solid.}\label{packing}
\end{figure}

\begin{table*}[h!]
\begin{tabular}{| c | c | c || c  | c | c |}
\hline
\; Regime \; & \; Relation \; & \; Prediction \; & Experiment & Frictionless Sim'n & \; Frictional Sim'n \; \\
\hline \hline
 &  $\delta \mu \sim \Ic^{\alpha_\mu}$ & $\alpha_\mu = 0.35$  & 1 \cite{Forterre08} & 0.38(4) \cite{Peyneau08} & \parbox[c][1cm]{3cm}{ 0.81(3) \cite{Peyneau09}, 1 \cite{Bouzid13}, \\ 1 \cite{Azema14}, 1 \cite{Trulsson12} } \\   
Inertial & $\delta \phi \sim \Ic^{\alpha_\phi}$ & $\alpha_\phi = 0.35$   & 1 \cite{Forterre08} & 0.39(1) \cite{Peyneau08}  & \parbox[c][2cm]{3cm}{0.87(2) \cite{Peyneau09}, 1 \cite{Azema14}, \\ 1 \cite{Trulsson12}} \\  
 & $\Lc \sim \Ic^{-1/2}$ & $1/2$ &  0.33 \cite{Menon97}, 0.3 \cite{Pouliquen04} & 0.48 \cite{Peyneau08}, 0.5 \cite{DeGiuli16} & 0.5 \cite{MiDi04}, 0.2 \cite{DeGiuli16} \\ 
 & $\epsilon_v \sim \Ic $ & 1 & 1 \cite{Menon97} & 1.1 \cite{DeGiuli16} & 0.95 \cite{DeGiuli16} \\
\hline
  & $\eta \sim |\delta \phi|^{-1/\gamma_\phi}$ & $\gamma_\phi^{-1} = 2.83$ & 2 \cite{Boyer11}, 2 \cite{Ovarlez06}&  \parbox[c][2cm]{4cm}{2.6(1) \cite{Olsson12}, 2.77(20) \cite{Olsson11},  \\ 2.56\cite{Kawasaki15},  2.77 \cite{DeGiuli15a} } & 1.5 \cite{Trulsson16} \\  
Viscous & $\delta \mu \sim \Jc^{\gamma_\mu}$ & $\gamma_\mu = 0.35$ & \parbox[c][1cm]{2.5cm}{0.38 \cite{Lespiat11}, 0.5 \cite{Boyer11} \\ 0.42 \cite{Cassar05,Lespiat11} } & \parbox[c][2cm]{4cm}{0.37 \cite{Peyneau09},  0.25 \cite{Olsson11}, \\ 0.32 \cite{DeGiuli15a} } & 0.5 \cite{Trulsson12} \\  
& $\delta z\sim \Jc^{\gamma_z}$ &$\gamma_z=0.30$ & &0.30\cite{DeGiuli15a}& \\
 & $\ell_c \sim |\delta \phi|^{-\gamma_\ell/\gamma_\phi}$ & $\gamma_\ell/\gamma_\phi = 0.43$ & & 0.6(1) \cite{Olsson07} & \\  
& $\Lc \sim \Jc^{-1/2}$ & $1/2$ &  & 0.5\cite{DeGiuli15a}, 0.5 \cite{Trulsson16} & 0.33 \cite{Trulsson16} \\
 & $\epsilon_v \sim \Lc^{-2} \sim \Jc $ & (-2,1) &  & $\epsilon_v \sim \Lc^{-2}$\cite{DeGiuli15a}, $\epsilon_v \sim \Jc$ \cite{Olsson10a} & \\
 & $d\Lc/d\epsilon \sim -\Lc^{3}$ & $3$ &  & 3 \cite{DeGiuli15a}& \\
& $\edot_{v\rightarrow i} \sim \delta \phi^{\gamma_{\edot}}$ & $\gamma_{\edot} = 2.83$ & 1 \cite{Fall10} & & \\
\hline
\end{tabular}
\caption{Predicted critical exponents {\it vs.} values from experiments and numerical simulations, with and without frictional interactions. The values extracted in Ref.\cite{Peyneau09} correspond to simulations closest to hard spheres (the ``roughness parameter" of that reference is $10^{-4}$). When available, error bars are indicated by the notation $0.38(4) = 0.38 \pm 0.04$, $2.77(20)=2.77 \pm 0.20$, etc. }
\end{table*}

Empirical values found for the exponents in Eqs.(\ref{0}-\ref{06}) are reported in Table 1. They seem not to depend on dimension, which we thus did not report in our table. In the case of inertial flow they appear to depend on the presence of friction, whereas for suspended particles exponents appear to be similar with and without friction. In this work we focus initially on theory for frictionless particles, and in a second section, discuss the effect of friction. 

Currently there is no accepted microscopic theory describing quantitatively these singular behaviors, in particular Eqs.(\ref{0}-\ref{06}). Observations support that as jamming is approached, particles form an extended network of contacts, and that the stress is dominated by contact forces \cite{Cruz05, Peyneau08,Boyer11}. In this work we review a framework to describe flow in such situations. The detailed arguments are presented in \cite{DeGiuli15a}; here we discuss the essential ingredients of the theory, and focus on the effect of friction, discussed in more detail in \cite{DeGiuli16,Trulsson16}.

We attack the problem  in two steps. First, we isolate the  microscopic quantities that control flow. Then, we compute the scaling properties of these quantities  by performing a perturbation around the solid phase. The idea is to consider the solid in the critical state, i.e. carrying the maximal stress ratio possible $\mu=\mu_c$, corresponding to a packing fraction $\phi=\phi_c$. Next, one adds an additional kick to the system, corresponding to a small  additional stress ratio $\delta \mu$. As a result, some contacts between particles will open, forces will be unbalanced, and the system will start to flow (see Fig. \ref{packing}).  Our key assumption is  that flowing configurations are similar to a solid that is thus destabilized. This approach  enables us to propose a full scaling description of the problem, and to predict the exponents entering Eqs.(\ref{0}-\ref{06}) in good agreement with observations  in the absence of static friction. Moreover, our approach predicts several other properties singular near jamming: the speed of the particles, the strain scale of velocity decorrelation, and the coordination of the contact network. The first two quantities are accessible experimentally, and provide an additional experimental test of our views. 

\begin{figure}[th!] 
\includegraphics[viewport = 30 75 755 430, width=\columnwidth,clip]{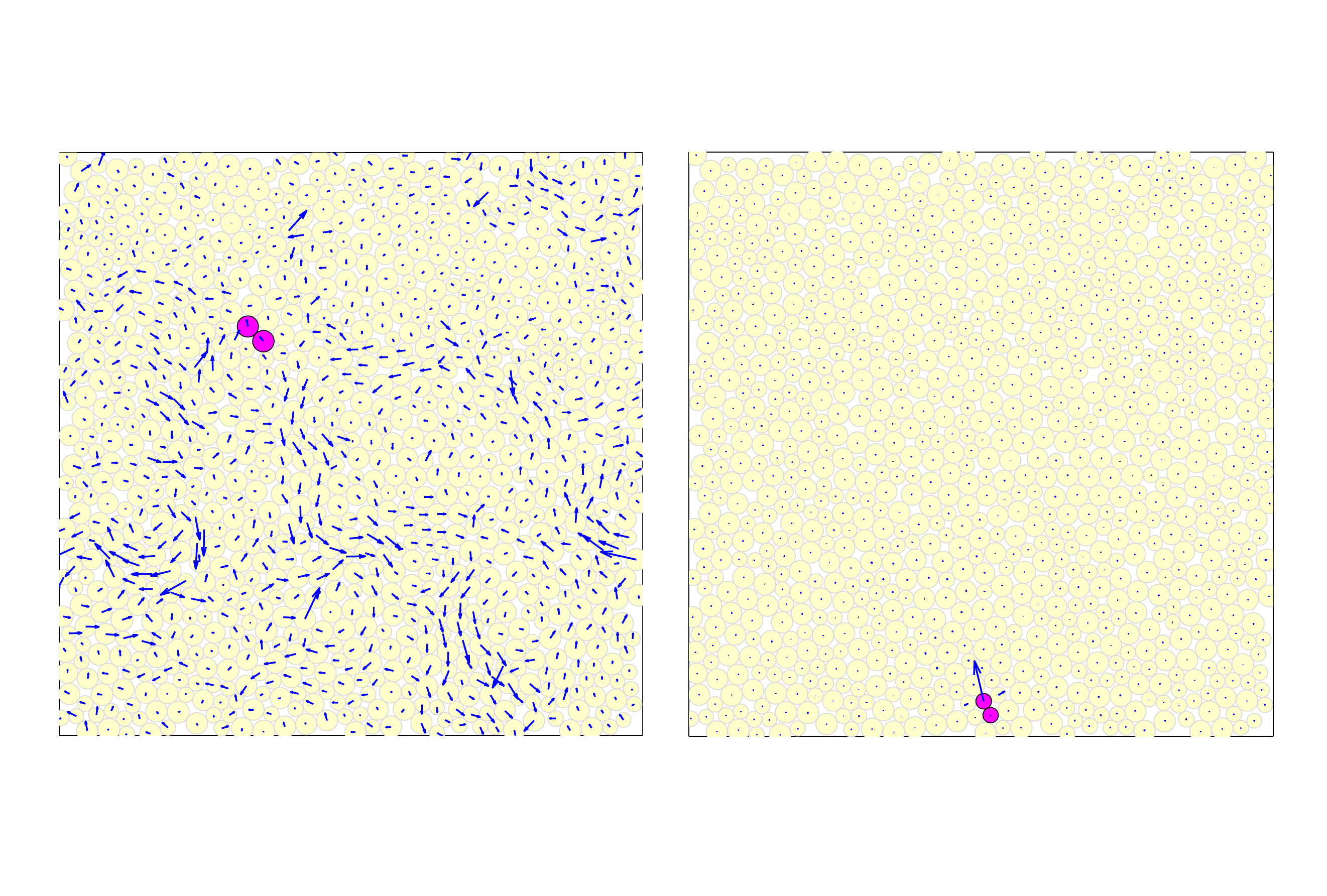} 
\caption{Extended {\it vs} localized contacts. When a contact is opened from an isostatic packing, the resulting deformation (arrows) can either be {\it extended}, as shown at left, or {\it localized}, as shown at right. Localized contacts are more numerous, but only extended contacts couple strongly to an imposed shear stress. Reproduced from \cite{Lerner13a} by permission of The Royal Society of Chemistry (RSC). }\label{twomodes}
\end{figure}

When an additional stress ratio is imposed to a marginal solid, naively one expects the contacts carrying the smallest forces to open. However, only a vanishingly small fraction of weak contacts are significantly coupled to external stresses \cite{Lerner13a}. We call them {\it extended} contacts, because perturbing such contacts mechanically lead to a spatially extended response in the system, as shown in Figure \ref{twomodes} \cite{Lerner13a,DeGiuli14b}. In a packing only those contacts lead to plasticity when stress is increased, or when a shock (say a collision) occurs in the bulk of the material  \cite{During16,Lerner13a}. The density of extended contacts as a function of the force $f$ in the contact follows
 \be
 \label{08}
 P(f)\propto \frac{1}{p^{1+\theta_e}} \; f^{\theta_e}.
 \ee
Numerically it is found that $\theta_e\approx 0.44$ both in two and three dimensions \cite{Lerner13a,DeGiuli14b}, suggesting that this quantity may be independent of dimension. 

Moreover, its value does not depend on the preparation protocol of the isostatic state: up to error bars, equal values are found from compression of hard spheres \cite{Lerner13a}, shear-jammed hard disks \cite{Lerner12}, and decompression of soft spheres \cite{DeGiuli14b,Charbonneau15} \footnote{We include here works where $\theta_e$ was not directly measured, but inferred from the force distribution exponent $\theta_\ell$ by the marginal stability relation $\theta_e=2\theta_\ell/(1-\theta_\ell)$. See \cite{DeGiuli14b}.}. The exponent $\theta_e$ can be shown to control the stability of the solid phase \cite{Wyart12,Lerner13a,Muller14}. Recently replica calculations in infinite dimension on the force distribution \cite{Charbonneau14,Charbonneau14a,Charbonneau15} led to the prediction \cite{DeGiuli14b}:
\be
\theta_e=0.423...,
\ee  
within the error bar of our measurements. In our proposed scaling description all exponents can be expressed in terms of $\theta_e$, in particular:
\ba
\alpha_\mu&=&\alpha_\phi=\gamma_\mu=\gamma_\phi= \frac{3+\theta_e}{8+4\theta_e}\approx 0.35 \\
\gamma_{\edot} &=& \frac{8+4\theta_e}{3+\theta_e} \approx 2.83 \\
\alpha_\ell&=& \gamma_\ell=   \frac{1+\theta_e}{8+4\theta_e}\approx 0.15  
\ea
Empirically it was noticed that $\gamma_\mu=\gamma_\phi$ and that $\alpha_\mu=\alpha_\phi$,
which our arguments rationalize. 

\section{Theory for Frictionless Particles}
\subsection{General Approach}
We argue that several dimensionless quantities that characterize the microscopic dynamics under flow critically affect rheological properties. 
As jamming is approached, the assembly of particles acts as a lever: due to steric hindrance, the typical relative velocity between adjacent particles $V_r$ becomes much larger than the characteristic velocity $\dot \epsilon D$ where  $\dot \epsilon$ is the strain rate and $D$ the mean radius of the particles \cite{Lerner12a,Andreotti12}. We thus define the amplitude of this lever effect $\Lc$ as:
\be
\label{1}
\Lc=\frac{V_r}{\dot \epsilon D}.
\ee
Another fundamental quantity, particularly relevant for inertial flow, is the strain scale $\epsilon_v$ beyond which a particle loses memory of its direction relative to its neighbors. $\epsilon_v$ can be extracted from the decay of the  autocorrelation function $\langle V^\alpha_r(0) V^\alpha_r(\epsilon)\rangle$, where the average is made over all pairs of adjacent particles $\alpha$. As the packing fraction $\phi$ increases toward jamming, collisions are more frequent per unit strain (due to the increase of relative particle motion $\Lc$), and each collision affects the motion of the particles on a growing length scale. These two effects imply that $\epsilon_v$ vanishes rapidly near jamming.

We now argue that dissipation is entirely governed by $\Lc$ in overdamped suspensions, and by both $\Lc$ and $\epsilon_v$ in inertial flows. In both cases the power injected into the system at the boundaries, which is simply ${\cal P}=\Omega \sigma \dot\epsilon$ at constant volume, must be dissipated in the bulk.  

In a dense suspension we expect dissipation to be governed by local mechanisms such as lubrication. Lubrication forces are singular for the ideal case of perfectly smooth spheres, but not for rough particles where they must be cut off. Thus the viscous force exchanged by two neighboring particles  must dimensionally follow $F\sim\eta_0 V_r D^{d-2}$, leading to a power dissipated ${\cal P}/N=C \eta_0 V_r^2 D^{d-2}$  where $d$ is the spatial dimension, $C$ is a dimensionless constant that depends on the particle shape and roughness, and $N$ is the number of particles. Equating the power dissipated to the power injected, one gets that for a given choice of particles:
\be
\label{2}
\frac{\eta}{ \eta_0}\propto 1/\Jc\propto \Lc^2
\ee
implying that the divergence of viscosity  is governed by  $\Lc$.  This  result holds by construction in simple models of dissipation in suspension flows \cite{Lerner12a,Andreotti12,Vagberg14b}.

Concerning inertial flows, we  suppose that the restitution coefficient characterizing a collision between two particles is smaller than one, and that collisions dominate dissipation. Then each time two neighboring particles change relative direction, a finite fraction of their relative kinetic energy $E_c\sim M V_r^2$  must be dissipated, where $M$ is the particle mass. Then the total power dissipated must follow ${\cal P}\propto N \dot \epsilon E_c /\epsilon_v$. Using Eq.(\ref{1}) and balancing power injected and dissipated, one gets $\sigma/(\dot\epsilon^2 D^2 \rho )\sim \Lc^2/\epsilon_v$ where $\rho$ is the mass density of the particles, so that the inertial number $\Ic$ follows:
\be
\label{4}
\Ic\sim \frac{\sqrt{\epsilon_v}}{\Lc}
\ee
To our knowledge Eq.(\ref{4}) has not been proposed before, and could be tested empirically. 

\subsection{Perturbation around the solid}
To obtain a complete description of flow, one must therefore express $\Lc$ and $\epsilon_v$ in terms of control parameters such as $\delta \mu$ or $\delta \phi$.  To achieve this goal, we make the assumption that the contact network of configurations in flow is similar to that of jammed configurations at $\mu_c$ immediately after increasing the stress ratio by $\delta \mu>0$. The detailed arguments in this section are presented in  \cite{DeGiuli15a}; here we discuss the physical ingredients of each argument and summarize the results.

The coordination $z$ of the network of contacts is a key microscopic quantity that distinguishes flowing from jammed configurations.  At jamming the coordination is just sufficient to forbid motion, corresponding to $z_c=2d$ for frictionless spheres \cite{Tkachenko99,Moukarzel98,Roux00}.  As illustrated in Fig.(\ref{packing}), the kick of amplitude $\delta \mu$ opens a fraction $\delta z\equiv z_c-z$ of the contacts, allowing collective motions of the particles for which particles do not overlap, but simply stay in contact, the so-called called {\it floppy modes}. Using a virtual work theorem to compute the work done after a contact is opened, we find \cite{DeGiuli15a} that the lever amplitude is directly related to the density of floppy modes $\delta z$, in particular
\be
\label{001}
\Lc\sim \delta z^{-(2+\theta_e)/(1+\theta_e)}
\ee
To see how many contacts are opened when an increment of stress ratio $\delta \mu$ is added to an isostatic packing, we can use the known behavior of the shear modulus near jamming. This results in 
\be
\label{002}
\delta z \sim \delta \mu^{(2+2\theta_e)/(3+\theta_e)}
\ee
Jointly Eqs.(\ref{001},\ref{002}) predict a relationship between lever amplitude and stress ratio: 
\be
\label{6}
\Lc\sim  \delta \mu^{-(4+2\theta_e)/(3+\theta_e)}\sim\delta\mu^{-1.41}
\ee 
Eqs.(\ref{2},\ref{6}) lead to a prediction for the exponent $\gamma_\mu$ entering in the constitutive relation $\mu(\Jc)$. 
Together with previous results showing that $\ell_c\sim 1/\sqrt{\delta z}$ \cite{During13,During14}, we obtain expressions for $\gamma_\ell$ and $\alpha_\ell$,
corresponding to:
\be
\label{6bis}
\ell_c\sim \delta \mu^{-(1+\theta_e)/(3+\theta_e)}\sim \delta\mu^{-0.41}
\ee
both for inertial and viscous flows. 

\begin{figure}[b!] 
 \centering
\includegraphics[width=0.8\columnwidth,clip]{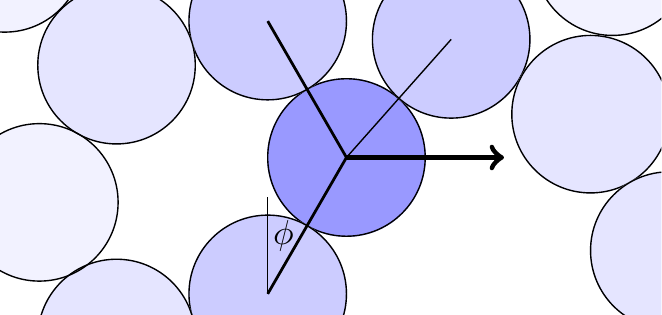}
\caption{Illustration of geometrical nonlinearity. If the central grain has an unbalanced contact force as indicated by the arrow, then the ensuing flow will tend to align the contact normals of the dominant contact forces (thick lines), i.e. the angle $\phi$ will increase. Geometrically, $d\phi/dt \propto V$, the velocity of the particle. }\label{fig_sketch}
\end{figure}

We now consider the characteristic strain scale $\epsilon_v$ at which velocities decorrelate. This is fixed by a geometrical argument: we use the fact that in dense flows, when a grain has an unbalanced net contact force, $\Fv$, the ensuing motion will tend to make the remaining contacts of the grain align along $\Fv$ (see Fig.\ref{fig_sketch}). Since forces are repulsive, this further increases the unbalanced contact force. The increase in force is proportional to the typical contact force, $p D^{d-1}$, as well as to the rotation of the contacts, of magnitude $\sim \Lc d\epsilon$, thus 
\eq{ \label{dF}
\frac{dF}{d\epsilon} \sim p D^{d-1}  \Lc,
}
where $\Lc$ is the dimensionless magnitude of the velocity fluctuation. This equation can also be derived formally, see \cite{DeGiuli15a}. In inertial flow, unbalanced forces are proportional to accelerations, $F = p \Ic^2 d\Lc/d\epsilon$, which leads to
\eq{ \label{epsv2}
\frac{d^2 \Lc}{d\epsilon^2} \propto \frac{\Lc}{\Ic^2}.
}
Eq.\ref{epsv2} indicates that there is a characteristic strain scale $\epsilon_v \approx \Ic$ in which a velocity fluctuation grows by an amount proportional to its initial magnitude. In steady flow, such growth must be destroyed by collisions on the same strain scale, since the latter reorganize the direction of particle motion. Hence $\epsilon_v$ is indeed the scale of decorrelation of particle velocities. Together with Eqs.(\ref{4},\ref{6}) this result leads to a prediction for the exponent $\alpha_\mu$ characterizing the constitutive relation $\mu(\Ic)$.

For viscous flows, we still have Eq.(\ref{dF}), but now unbalanced contact forces are equal to the viscous drag forces, since the total net force vanishes. In this case $\Fv_i \approx \eta_0 \Vv_i^{na}$, where ${}^{na}$ indicates the non-affine part. Assuming $\Vv_i^{na} \approx V_r$, we get $F/p \approx (\eta_0 \edot/p) \Lc = \Lc\Jc$, leading to 
\eq{ \label{epsv3}
\frac{d\Lc}{d\epsilon} \propto \frac{\Lc}{\Jc}. 
}
By a similar argument as above, this implies a characteristic strain scale $\epsilon_v \approx \Jc$ in which velocity fluctuations are created and destroyed.

One missing link to obtain a full scaling description of the problem is how the packing fraction depends on other control parameters. If we make the additional assumption that isotropic packings of frictionless particles in the thermodynamic limit have a finite (although presumably small) dilatancy, then this implies the scaling relation $\delta\phi\sim \delta \mu$, known to agree well with observations \cite{DeGiuli15a}. This result enables us to predict the exponents $\alpha_\phi$ and $\gamma_\phi$ entering the constitutive relation for $\phi(\Ic)$ and $\phi(\Jc)$, leading to a complete scaling description of rheological properties near jamming for frictionless particles. In particular the divergence of viscosity with packing fraction in suspensions is expected to follow:
\be
\label{8bis}
\frac{\eta}{\eta_0}\sim (\phi_c-\phi)^{-(8+4\theta_e)/(3+\theta_e)}\sim (\phi_c-\phi)^{-2.83}
\ee
Finally, by a straightforward analysis one can extract the strain rate when one expects a transition from viscous to inertial flow, leading to Eq.(\ref{epsco}) \cite{DeGiuli15a}. 

Our results are compared to previous empirical and numerical observations in Table 1. Overall we find a very good agreement between observations and predictions for frictionless particles. However, friction appears to change exponents for inertial flows, and may also affect viscous flows. 

\begin{figure}[t!] 
\includegraphics[width=0.48\textwidth,viewport=15 5 230 100,clip]{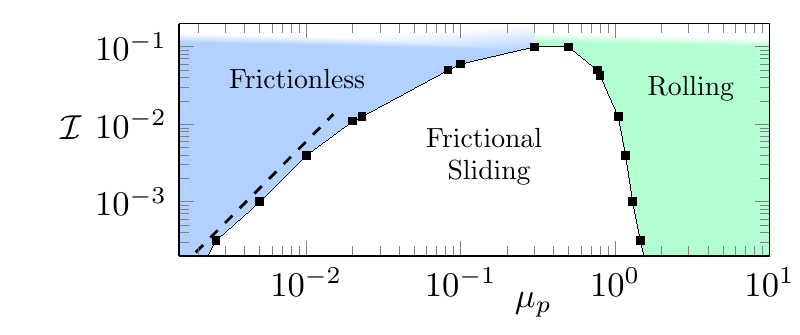}
\caption{Phase diagram of dense homogeneous inertial frictional flow. In the frictionless and rolling regimes, most energy is dissipated by inelastic collisions, while in the frictional sliding regime energy dissipation is dominated by sliding. Along the phase boundary, grains dissipate equal amounts of energy in collisions and in sliding. For $\Ic \gtrsim 0.1$, one enters the dilute regime \cite{Azema14}. The dashed line has slope $2$.\label{phase}}
\end{figure}

\section{Effect of Friction}
\subsection{Phase Diagram}
In the theory for frictionless particles, the fact that the power injected and the power dissipated exactly compensate on average plays a crucial role in relating the strain rate to geometrical quantities, as expressed in Eqs.\ref{2},\ref{4}. Central to the arguments are the dominant dissipation mechanisms, assumed above to be viscous dissipation in over-damped  dynamics, and grain inelasticity in inertial dynamics. However, when particles are frictional, energy will also be dissipated by sliding at frictional contacts, in addition to the other dissipation mechanisms  above. As Coulomb friction coefficient $\mu_p$ and shear rate are varied, it is not obvious {\it a priori} in which regimes the sliding dissipation rate $\Dc_{slid}$ will dominate over inelastic collisions in inertial flows, and viscous dissipation in viscous flows; we call these latter sources of dissipation $\Dc_0$.

\begin{figure}[t!]
\hspace*{-0.7cm}\includegraphics[width=1.1\columnwidth]{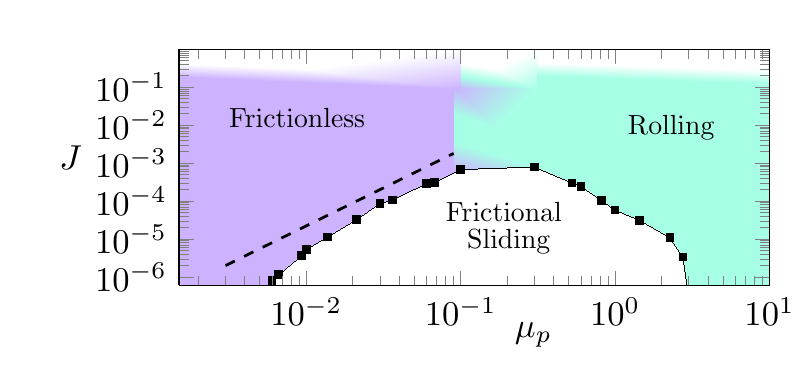}
\caption{Phase diagram of dense non-Brownian suspension flow. In the Frictionless and Rolling regimes, the dominant source of dissipation is viscous drag, whereas in the Frictional Sliding regime, dissipation is dominated by sliding friction. The dashed line has slope 2.}
\label{phasevisc}
\end{figure}

This question can be straightforwardly investigated with numerical simulations. To map out the phase diagram as $\mu_p$ and shear rate are varied, we considered, separately, the parameter space $(\mu_p, \Ic)$ for inertial dynamics \cite{DeGiuli16}, and $(\mu_p, \Jc)$  for viscous dynamics \cite{Trulsson16}. The main results are the phase diagrams presented in Figs.\ref{phase},\ref{phasevisc}. For both viscous and inertial dynamics, we find 3 phases: (i) a `Frictionless' regime in which $\Dc_0$ dominates, most contacts are sliding, and investigated quantities are consistent with the theory for frictionless particles, in particular $\Lc$, $\epsilon_v$, $\delta \mu$, and $\delta \phi$; (ii) a `Rolling' regime in which again $\Dc_0$ dominates, but most contacts are rolling, not sliding. Here $\Lc$, $\delta \mu$, and $\delta \phi$ appear to be consistent with the frictionless theory, particularly in the limit $\mu_p \gg 1$. Finally, in between these regimes we find (iii) a `Frictional Sliding' regime in which $\Dc_{slid} > \Dc_0$. Here $\epsilon_v \sim \Ic (\Jc)$ in inertial (viscous) dynamics, as predicted by Eqs.\ref{dF},\ref{epsv2},\ref{epsv3}, which hold in the presence of friction, but other quantities have new scalings: for example, $\Lc$ has a weaker divergence, and $\delta \mu$ has a much less dramatic behavior at small strain rate. 

The crossover from `Frictionless' to `Frictional Sliding' regimes can be predicted from the theory above. Indeed, $\Dc_{slid}$ is simply the sum over sliding contacts of $\fv_\alpha{}^{\! T} \cdot \Uv_\alpha^{T}$, where $T$ indicates the tangential component. Denoting by $\chi$ the fraction of contacts that are sliding, we have $\Dc_{slid} \approx N_C \chi \mu_p \langle f_N \rangle \Lc \edot$, where $N_C$ is the total number of contacts and $\langle f_N \rangle \sim p$ is the typical normal force. Inside the frictionless regime, nearly all contacts are sliding, $\chi \approx 1$. Therefore we find that sliding friction constitutes a fraction 
\eq{
\frac{\Dc_{slid}}{\Pc} \sim \frac{N \mu_p p \edot \Lc}{\Omega \sigma \edot} \sim \frac{\mu_p}{\mu_c} \Lc 
}
of the total energy dissipation rate (which must equal $\Pc$ in steady flow). Since $\Lc$ is diverging as jamming is approached (Eq.\ref{6}) we thus expect a transition to a regime dominated by frictional sliding, both for inertial and viscous flows, as observed. From Eqs.(\ref{2},\ref{4}) and $\epsilon_v \sim \Ic$, it follows that $\Lc \sim \Ic^{-1/2}$ and $\Lc \sim \Jc^{-1/2}$ in frictionless inertial and viscous dynamics, respectively, thus leading to a crossover `Frictionless' to `Frictional Sliding' at $\Ic \sim \mu_p^2$ and $\Jc \sim \mu_p^2$. These scalings are plotted as dashed lines in Figs.\ref{phase},\ref{phasevisc}, and are consistent with the data. 

\subsection{Frictional Theory}
Still missing is a complete theory of the Frictional Sliding regime. We expect Eqs.\ref{dF},\ref{epsv2},\ref{epsv3} to be always valid, and indeed the associated prediction $\epsilon_v \sim \Ic$ appears to hold in the Frictional Sliding regime as well. However, the velocity fluctuations and rheological properties present new scalings. In particular, in all cases the velocity fluctuations are large and their spatial correlations are long-range as jamming is approached. For frictionless particles, a single characteristic scale characterizes these velocity fluctuations. By contrast, in \cite{Trulsson16} we provided evidence that several scales are required to describe the kinetics of frictional particles. For example, the relative velocity between contacting particles differs from the amplitude of non-affine velocities. This additional complexity is plausibly caused by the intermittent and non-extensive localization of strain observed in several studies \cite{Heussinger13,Henkes16,Maiti16,Kharel16}.  A similar localization is observed in the plasticity of soft particles \cite{Maloney06}, whose connection to  hard frictional particles is, however, unclear. 
A possible factor causing the difference between frictional and frictionless particles relates to the criticality of the marginal solid: for hard frictionless spheres, Maxwell counting leads to a precise identification of the marginal state, which is exactly isostatic. By contrast,  frictional solids at jamming are generally hyperstatic \cite{Wyart05,Liu10,Hecke10}. How these issues are related are outstanding problems. 

\section{Conclusion}

We reviewed a theory for flow of hard, frictionless particles, both in viscous and inertial dynamics. In a first step, we related the power dissipated in flow of frictionless particles to certain microscopic kinetic quantities. The latter control singularities in the rheological properties near jamming. In a second step, we have computed these quantities, using a perturbation around the solid phase. Our main hypothesis is that
configurations in flow are similar to jammed configurations at maximum stress ratio $\mu_c$, destabilized by an additional stress increment $\delta \mu$. In this approach, the properties of the solid phase are central, in particular the fact that  the density of contacts which can couple to external forces is singular at small forces, and characterized by a non-trivial exponent $\theta_e$. Our description of flow can thus be thought as that of a jammed solid, populated by elementary excitations corresponding to the opening of weak contacts, of density $\delta z$. 

When friction is added, the theory holds in a finite region of parameter space, whose boundary is well predicted by the theory. Theory for the regime in which sliding friction dominates is, however, lacking. In our view, a central question for the future is what controls the stability of isostatic frictional systems, how these respond to an additional stress ratio $\delta \mu$, and how the combination of finite softness, inertia and friction qualitatively affects the flow curves \cite{Otsuki11}.

\section*{Acknowledgements and References}
We are pleased to thank our collaborators on this project: Gustavo D\"uring, Edan Lerner, Jim McElwaine, and Martin Trulsson. 

%
 \bibliography{../bib/Wyartbibnew}

\begin{thebibliography}{52}

\bibitem{MiDi04}
G.~MiDi, The European Physical Journal E: Soft Matter and Biological Physics
  \textbf{14}, 341 (2004-08-01)

\bibitem{Cruz05}
F.~da~Cruz, S.~Emam, M.~Prochnow, J.N. Roux, F.m.c. Chevoir, Phys. Rev. E
  \textbf{72}, 021309 (2005)

\bibitem{Jop06}
P.~Jop, Y.~Forterre, O.~Pouliquen, Nature \textbf{441}, 727 (2006)

\bibitem{Pouliquen04}
O.~Pouliquen, Physical review letters \textbf{93}, 248001 (2004)

\bibitem{Olsson07}
P.~{Olsson}, S.~{Teitel}, Phys.\ Rev.\ Lett. \textbf{99}, 178001 (2007)

\bibitem{During14}
G.~D{\"u}ring, E.~Lerner, M.~Wyart, Physical Review E \textbf{89}, 022305
  (2014)

\bibitem{Lemaitre09b}
A.~Lema{\^\i}tre, J.N. Roux, F.~Chevoir, Rheologica acta \textbf{48}, 925
  (2009)

\bibitem{Boyer11}
F.~Boyer, E.~Guazzelli, O.~Pouliquen, Phys. Rev. Lett. \textbf{107}, 188301
  (2011)

\bibitem{Fall10}
A.~Fall, A.~Lemaitre, F.~Bertrand, D.~Bonn, G.~Ovarlez, Phys. Rev. Lett.
  \textbf{105}, 268303 (2010)

\bibitem{Trulsson12}
M.~Trulsson, B.~Andreotti, P.~Claudin, Physical review letters \textbf{109},
  118305 (2012)

\bibitem{Vagberg14}
D.~V{\aa}gberg, P.~Olsson, S.~Teitel, Physical Review Letters \textbf{112},
  208303 (2014)

\bibitem{Forterre08}
Y.~Forterre, O.~Pouliquen, Annual Review of Fluid Mechanics \textbf{40}, 1
  (2008)

\bibitem{Peyneau08}
P.E. Peyneau, J.N. Roux, Physical review E \textbf{78}, 011307 (2008)

\bibitem{Peyneau09}
P.E. Peyneau, Ph.D. thesis, Ecole des Ponts ParisTech ({2009})

\bibitem{Bouzid13}
M.~Bouzid, M.~Trulsson, P.~Claudin, E.~Cl{\'e}ment, B.~Andreotti, Phys. Rev.
  Lett. \textbf{111}, 238301 (2013)

\bibitem{Azema14}
E.~Az{\'e}ma, F.~Radjai, Physical review letters \textbf{112}, 078001 (2014)

\bibitem{Menon97}
N.~Menon, D.J. Durian, Science \textbf{275}, 1920 (1997)

\bibitem{DeGiuli16}
E.~DeGiuli, J.~McElwaine, M.~Wyart, Phys. Rev. E \textbf{94}, 012904 (2016)

\bibitem{Ovarlez06}
G.~Ovarlez, F.~Bertrand, S.~Rodts, Journal of rheology \textbf{50}, 259 (2006)

\bibitem{Olsson12}
P.~Olsson, S.~Teitel, Physical review letters \textbf{109}, 108001 (2012)

\bibitem{Olsson11}
P.~Olsson, S.~Teitel, Physical Review E \textbf{83}, 030302 (2011)

\bibitem{Kawasaki15}
T.~Kawasaki, D.~Coslovich, A.~Ikeda, L.~Berthier, Physical Review E
  \textbf{91}, 012203 (2015)

\bibitem{DeGiuli15a}
E.~DeGiuli, G.~D\"uring, E.~Lerner, M.~Wyart, Physical Review E \textbf{91},
  062206 (2015)

\bibitem{Trulsson16}
M.~Trulsson, E.~DeGiuli, M.~Wyart, arXiv preprint arXiv:1606.07650  (2016)

\bibitem{Lespiat11}
R.~Lespiat, S.~Cohen-Addad, R.~H\"ohler, Phys. Rev. Lett. \textbf{106}, 148302
  (2011)

\bibitem{Cassar05}
C.~Cassar, M.~Nicolas, O.~Pouliquen, Physics of Fluids \textbf{17}, 103301
  (2005)

\bibitem{Olsson10a}
P.~Olsson, Phys. Rev. E \textbf{81}, 040301 (2010)

\bibitem{Lerner13a}
E.~Lerner, G.~During, M.~Wyart, Soft Matter \textbf{9}, 8252 (2013)

\bibitem{DeGiuli14b}
E.~DeGiuli, E.~Lerner, C.~Brito, M.~Wyart, Proceedings of the National Academy
  of Sciences \textbf{111}, 17054 (2014)

\bibitem{During16}
G.~D{\"u}ring, E.~Lerner, M.~Wyart, arXiv preprint arXiv:1602.08317  (2016)

\bibitem{Lerner12}
E.~Lerner, G.~D\"uring, M.~Wyart, EPL (Europhysics Letters) \textbf{99}, 58003
  (2012)

\bibitem{Charbonneau15}
P.~Charbonneau, E.I. Corwin, G.~Parisi, F.~Zamponi, Physical Review Letters
  \textbf{114}, 125504 (2015)

\bibitem{Wyart12}
M.~Wyart, Phys. Rev. Lett. \textbf{109}, 125502 (2012)

\bibitem{Muller14}
M.~M{\"u}ller, M.~Wyart, Annual Review of Condensed Matter Physics \textbf{6},
  177 (2015)

\bibitem{Charbonneau14}
P.~Charbonneau, J.~Kurchan, G.~Parisi, P.~Urbani, F.~Zamponi, Nature
  communications \textbf{5} (2014)

\bibitem{Charbonneau14a}
P.~Charbonneau, J.~Kurchan, G.~Parisi, P.~Urbani, F.~Zamponi, Journal of
  Statistical Mechanics: Theory and Experiment \textbf{2014}, 10009 (2014)

\bibitem{Lerner12a}
E.~Lerner, G.~D\"uring, M.~Wyart, Proceedings of the National Academy of
  Sciences \textbf{109}, 4798 (2012)

\bibitem{Andreotti12}
B.~Andreotti, J.L. Barrat, C.~Heussinger, Phys. Rev. Lett. \textbf{109}, 105901
  (2012)

\bibitem{Vagberg14b}
D.~V{\aa}gberg, P.~Olsson, S.~Teitel, Physical Review Letters \textbf{113},
  148002 (2014)

\bibitem{Tkachenko99}
A.V. Tkachenko, T.A. Witten, Phys. Rev. E \textbf{60}, 687 (1999)

\bibitem{Moukarzel98}
C.F. Moukarzel, Phys. Rev. Lett. \textbf{81}, 1634 (1998)

\bibitem{Roux00}
J.N. Roux, Phys. Rev. E \textbf{61}, 6802 (2000)

\bibitem{During13}
G.~D{\"u}ring, E.~Lerner, M.~Wyart, Soft Matter \textbf{9}, 146 (2013)

\bibitem{Heussinger13}
C.~Heussinger, Physical review E \textbf{88}, 050201 (2013)

\bibitem{Henkes16}
S.~Henkes, D.A. Quint, Y.~Fily, J.M. Schwarz, Physical Review Letters
  \textbf{116}, 028301 (2016)

\bibitem{Maiti16}
M.~Maiti, A.~Zippelius, C.~Heussinger, arXiv preprint arXiv:1606.06038  (2016)

\bibitem{Kharel16}
P.~Kharel, P.~Rognon, arXiv preprint arXiv:1605.00337  (2016)

\bibitem{Maloney06}
C.E. Maloney, Phys. Rev. Lett. \textbf{97}, 035503 (2006)

\bibitem{Wyart05}
M.~Wyart, S.~Nagel, T.~Witten, EPL (Europhysics Letters) \textbf{72}, 486
  (2005)

\bibitem{Liu10}
A.J. Liu, S.R. Nagel, W.~van Saarloos, M.~Wyart, \emph{The jamming scenario: an
  introduction and outlook} (Oxford University Press, Oxford, 2010)

\bibitem{Hecke10}
M.~van Hecke, Journal of Physics: Condensed Matter \textbf{22}, 033101 (2010)

\bibitem{Otsuki11}
M.~Otsuki, H.~Hayakawa, Physical Review E \textbf{83}, 051301 (2011)

\end{thebibliography}
%

\end{document}